\documentclass[twocolumn,showpacs,preprintnumbers,amsmath,amssymb]{revtex4}
\usepackage{graphicx}% Include figure files
\usepackage{dcolumn}% Align table columns on decimal point
\usepackage{bm}% bold math

\RequirePackage{cmap}
\RequirePackage[cp1251]{inputenc}
\newcommand{\frat}[2]{\frac{\textstyle #1}{\textstyle #2}}
\newcommand{\vf}[1]{\mbox{\boldmath $#1$}}
\newcommand{\nomer}[1]{\mbox{$\cal N$\hspace{-.5ex}\raisebox{.3ex}
           {\underline{\tiny 0}$\!$} #1}}

\begin{document}

\title{STUDYING QUARK CONDENSATES \\WITHIN MODELS OF FOUR-QUARK INTERACTIONS}

\author{S. V. Molodtsov}
 \altaffiliation[Also at ]{
Institute of Theoretical and Experimental Physics, Moscow, RUSSIA}
\affiliation{%
Joint Institute for Nuclear Research, Dubna,
Moscow region, RUSSIA%\textbackslash\textbackslash
}%
\author{G. M. Zinovjev}
\affiliation{
Bogolyubov Institute for Theoretical Physics,
National Academy of Sciences of Ukraine, Kiev, UKRAINE
}%

\date{\today}% It is always \today, today,

\begin{abstract}
Analysing two models of four-quark interactions which are of intrinsic difference in the
behaviours of their correlation lengths some
issues of quark condensations are considered. It is demonstrated that the quark
condensates substantially are not sensitive to the details of those interactions in the
range of coupling constants interesting for applications.
\end{abstract}

\maketitle

Studying the phase structure of strongly interacting matter is one of the most complicated tasks
of up-to-date theoretical development and any theoretical achievement here is highly non-trivial
being very indicative and guiding for planning and designing next round of physics experiments.
It concerns not only the heavy ion experiments which are very instrumental to analyse the phase
diagram of QCD matter at high temperatures and low baryon number densities but also the
fundamental astrophysical research as being very informative to trace the phase diagram at low
temperatures and very high baryon number densities \cite{ebw}. Moreover, nowadays it becomes
clear these sources of experimental information on matter phase structure should take into
account the strong magnetic fields that are generated in non-central collisions of relativistic
heavy ions and are present in cores of neutron stars. Magnetic fields induce
non-trivial effects in quark matter probing, first of all, the topological gluon configurations
at high temperatures, and modifying color superconducting phase at high baryon number densities
and low temperatures.

Obviously, the reliable information may be obtained only in these two extreme regions
(asymptotically high temperatures or baryon number densities) of the QCD phase diagram where
perturbative calculations are possible, and the studies at finite values of these parameters
rely on the model considerations (or effective field theories). As known a dynamical content of
relativistic quantum theories is ciphered in their correlation functions which are usually rather
complicated functions of incoming and outgoing particle momenta. If one realizes the source of
the most complicated behavior in a correlation function (for example, the standard propagators
of the intermidiate states) it becomes clear the contribution of heavy states to the correlation
function of interest will be well approximated by the first terms of series expansion if the
kinematics allows. Following such an idea in expanding the Lagrangian in powers of external
momenta of the light fields leads to an effective field theory which allows to proceed further
into a non-perturbative region with maintaining the symmetry constraints of the underlaying
theory \cite{polch}.

The investigation of color superconductivity problems \cite{blr} that followed such a strategy
provided us with a new look at the QCD ground state (a quark condensate) in a region of high
baryon number densities and shed a new light on the possible QCD phase structure. However, many
interesting questions of this research field are still unanswered and we try to clarify those in
current note. In particular, if a quark interaction is approximated as a point-like it is not
clear what takes place beyond the momentum cut-off and what is a true condensate profile as a
function of quark momentum. The nonlocal models with diverse ensembles of gluon vacuum fields are
dealing with a zero-mode approximation \cite{dc} but a pressing demand to get out of this
approach has been discussed (see, for example, \cite{ker}). It remains still unclear in which
extent a separable form of interaction, that is quite convenient for analytical study, distorts
a genuine picture as well as a role of quark and anti-quark condensates in the different
coupling (scalar-pseudoscalar or vector) channels \cite{ko}. Novel moment of our analysis is
based on the view of these questions from two akin but polar opposite (in a sense, which becomes
clear further) models.  Furthermore, our general conclusion about an approximate absence of the
interaction form influence on the dynamical characteristics of quark ensemble in a practical
interval of coupling constant \cite{MZ} makes a thorough analysis of the point rather actual.

Here Hamiltonian density which we are interested in has the form
\begin{equation}
%1
\label{1}
{\cal H}=-\bar q~(i{\vf \gamma}{\vf \nabla}+m)~q-j^a_\mu \int d{\vf y}~
\langle A^{a}_\mu A'^{b}_\nu\rangle~j'^b_\nu~,
\end{equation}
where $j^a_\mu=\bar qt^a\gamma_\mu q$ is the quark current, with operators of the quark fields
$q$, $\bar q$, taken in spatial point ${\vf x}$ (the variables with prime corresponds to the
${\vf y}$ point), $m$ is the current quark mass, $t^a=\lambda^a/2$ is the color gauge group
$SU(N_c)$ generators, $\mu,\nu=0,1,2,3$. The gluon field correlator
$\langle A^{a}_\mu A'^{b}_\nu\rangle$ is taken in the simplest color singlet form with a time
contact interaction (without retardation)
\begin{equation}
%2
\label{cor}
\langle A^{a}_\mu A'^{b}_\nu\rangle=G~\delta^{ab}~\delta_{\mu\nu}~F({\vf x}-{\vf y})~,
\end{equation}
(we do not include corresponding delta-function on time in this formula). Generally speaking
the terms spanned on the relative distance are permitted but for the sake of simplicity we
neglect corresponding contribution. This simple correlation function is a fragment of
corresponding ordered exponent and besides four-fermion interaction accompanied infinite number
of multi-fermion vertices arise. But for our purposes it would be quite enough to restrict
ourselves with this simple form. The mentioned above effective interactions appear in natural way
by the coarse-grained description of the system with the corresponding averaging procedure having
in mind that the vacuum gluon field changing stochastically (for example, in the form of
instanton liquid, see \cite{MZ}). A formfactor $F({\vf x})$ is interpreted in an simple manner
as an interaction potential of point-like particles. The correlation function itself looks like
formally a gauge non-invariant object, but it turns out that there exist an effective way to
compensate this shortage by examining, in some sense, all potentials which would be of interest.
For example, this set would be convincing enough if it is possible to compare two akin but
opposite potentials, one behaving as the delta function formfactor in the coordinate space
(one-flavour Nambu--Jona--Lasinio (NJL) model \cite{njl}) and another as the delta function
formfactor in the momentum space, an analog of which is well known in condensed matter physics
as the Keldysh model (KKB) \cite{kldsh} and corresponds to an infinite correlation length in the
coordinate space. It is worth of remarking here that the only feature of this model essential for
us further concerns the fact that the complicated integral equations become algebraic ones thanks
to the formfactor form. Tuning the scale of coupling constant $G$ that is interesting for
applications should be done by attaching it to the corresponding PDG meson observables.

It is believed that at sufficiently large interaction the ground state of the system transforms
from trivial vacuum $|0\rangle$ (the vacuum of free Hamiltonian) into the mixed state (with
quark--anti-quark pairs with the opposite momenta of vacuum quantum numbers) which is presented
as the Bogolyubov trial function (in that way some separate reference frame is introduced,
and chiral phase becomes fixed)
$$
|\sigma\rangle={\cal{T}}|0\rangle,~
{\cal{T}}=\prod\limits_{ p,s}\exp[\varphi_p~(a^+_{ p,s}b^+_{- p,s}+
a_{ p,s}b_{-p,s})].$$
Here $a^+$, $a$ and $b^+$, $b$ are the quark creation and annihilation operators, $a|0\rangle=0$,
$b|0\rangle=0$. The dressing transformation ${\cal{T}}$ transmutes the quark operators to the
creation and annihilation operators of quasiparticles
$A={\cal{T}}~a~{\cal{T}}^\dagger$, $B^+={\cal{T}}~b^+{\cal{T}}^\dagger$.

\begin{figure}%[!tbh]
\includegraphics[width=0.3\textwidth]{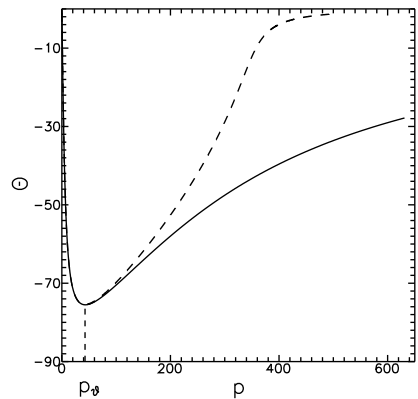}
\caption{The most stable equilibrium angles $\theta$ (in degrees)
as function of momentum $p$ in MeV. The solid line shows the result for the NJL model,
dashed one corresponds to the KKB model.
}
\label{f1}
\end{figure}

The pairing angle can be found from the condition of mean energy minimum
$\langle\sigma| H |\sigma\rangle$. Investigating the Bogolyubov transformation as a function of
formfactor demonstrates that the quark--anti-quark pairing angle (dynamical quark mass)
does not show any essential dependence on the formfactor profile \cite{MZ}. The most profitable
coupling angles $\theta=2\varphi$ are presented for comparison in Fig.\ref{f1} with the solid
line for the NJL model and dashed one for the KKB model under normal conditions ($T=0$, $\mu=0$).
For the delta-like potential in coordinate space (the NJL model) the mean energy diverges and to
obtain the reasonable results the upper limit cutoff in the momentum integration $\Lambda$ is
introduced being one of the tuning model parameters along with the coupling constant $G$ and
current quark mass $m$. Below we use one of the standard sets of the parameters for the NJL model
\cite{5}: $\Lambda=631$ MeV, $G\Lambda^2/(2\pi^2)\approx 1.3$, $m=5$ MeV, whereas the KKB model
parameters are chosen in such a way that for the same quark current masses the dynamical quark
ones in both NJL and KKB models coincide at vanishing quark momentum. It is interesting to notice
that the momentum $p_\vartheta$ (parameter) corresponds to the maximal attraction between quark
and anti-quark. The value of this parameter reversed determines a characteristic size of quasiparticle.
%%%%%%%%%%%%%%%%%%%%%%%%%%%%%%%%%%%%%%%%%%%%%%%%%
%%%%%%%%%%%%%%%%%%%%%%%%%%%%%%%%%%%%%%%%%
For the models under consideration it is of order of $p_\vartheta \sim(m M_q)^{1/2}$,
where $M_q$ is a characteristic quark dynamical mass, i.e. the quasiparticle
size is comparable with the size of $\pi$-meson (Goldstone particle). It is a
remarkable fact that the quasiparticle, as it is seen from Fig. \ref{f1}, does not depend
noticeably on the formfactor profile or, in other words, on the scale, but is rather
dependent on the coupling constant. It is worthwhile to mention here that now we understand
(with high clarity) the vacuum ensemble establishes a characteristic scale of the correlation
length order. Then the KKB model limit corresponds just to the characteristic system size. The
gluon correlation functions as measured in the lattice calculations show for the characteristic
size of corresponding configurations the estimates $0.1$--$0.2 fm$ \cite{cor1} (see also \cite{cor2}).
Besides, the lattice data for gluon propagator which could be interpreted as a gluon mass
generation in a finite momentum interval support such an estimate \cite{cor3}.

\begin{figure}%[!tbh]
\includegraphics[width=0.3\textwidth]{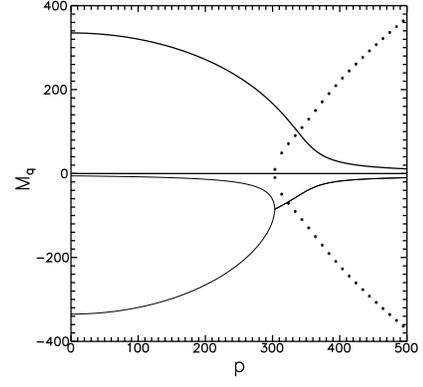}
\caption{Three branches of solutions for dynamical quark mass in MeV for the
KKB model as a function of momentum (MeV). Dots demonstrate the imaginary
part of the solutions.}
\label{f2}
\end{figure}

Dynamical quark mass $M_q$ can be expressed via a pairing angle by the relation
\begin{equation}
%3
\label{3}
\sin \left(\theta-\theta_m\right)=\frat{M_q}{P_0}~,
\end{equation}
where $P_0=[{\vf p}^2+M_q^2({\vf p})]^{1/2}$ is the energy of quark quasiparticle,
below notation $E_p$ would also be applied.
The auxiliary angle $\theta_m$ is determined by the relation $\sin \theta_m=m/p_0$,
where $p_0=[{\vf p}^2+m^2]^{1/2}$.
It can be shown that the dynamical quark mass can be determined through the equation
\begin{equation}
%4
\label{4}
M_q({\vf p})=m+2G\int d \widetilde{\vf q}~(1-n'-\bar n')~\frac{M'_q}{P'_0}~F({\vf p}+{\vf q}),
\end{equation}
where $n$, $\bar n$ is the quark antiquark distribution function under external conditions,
$\beta=T^{-1}$, $T$ is the ensemble temperature, $\mu$ is the quark chemical potential
\begin{equation}
%5
\label{5}
n=\left [e^{\beta(P_0-\mu)}+1\right]^{-1},
~\bar n=\left[e^{\beta(P_0+\mu)}+1\right]^{-1},
\end{equation}
integration is performed over momentum $\widetilde{\vf p}={\vf p}/(2\pi)^3$.
The relation between coupling constants $\widetilde g$ and $G$ would be considered below.
In particular at normal conditions ($T=0$, $\mu=0$) the dynamical quark mass in the NJL model
is $M_q\sim 340$ MeV. Dynamical quark mass in the KKB model ($F({\vf p})=\delta({\vf p})$) is
defined by the equation
\begin{equation}
%6
\label{6}
M({\vf p})=2 G~\frac{M_q({\vf p})}{P_0}~,
\end{equation}
and the dynamical quark mass is related to the induced quark mass by the relation $M_q=m+M$.
In practice, it turns out to be more convenient to use an inverse
function $p(M_q)$.
In particular, in the chiral limit  $M_q=(4G^2-{\vf p}^2)^{1/2}$, at $|{\vf p}|<2G$,
$M_q=0$, at $|{\vf p}|>2G$.
Then, the quark states with momenta $|{\vf p}|<2G$ are degenerate in energy $P_0=2G$.
Fig.\ref{f2} demonstrates three branches of the equation (\ref{6}) solutions for dynamical quark
mass. The dots show the imaginary part of solutions which are generated at the point where two
real solution branches are getting merged.
%%%%%%%%%%%%%%%%%%%%%%%%%%%%%%%%%%%%%%%%%%%%%%%%%
%%%%%%%%%%%%%%%%%%%%%%%%%%%%%%%%%%%%%%%%%

Now focusing on the analysis of the KKB model in mean field approximation
we present the Lagrangian density as follows:
\begin{equation}
%7
\label{7}
{\cal L}=\bar q~(i\gamma_\mu \partial_\mu+\hat \mu-m)~q +\widetilde g~j^a_\mu j'^a_\mu~,
\end{equation}
where additional summand  with the quark chemical $\hat \mu=\mu \gamma_0$ potential
was introduced for a convenient work with the Green’s function. We do not show here an
integration over ${\bf y}$ coordinate with the corresponding formfactor to simplify our setup
and recall only that the primed variables are associated with point ${\bf y}$.

Making some transformations with color matrices arising at the Fierz transformation
of the quark fields in mean field approximation and taking the well-known relations for color
SU(3) group generators
$$
{\vf \lambda}^i_j{\vf \lambda}^k_l=2~\delta^i_l\delta^k_j-
\frat{2}{3}\delta^i_j\delta^k_l~,$$
together with the identity
$$\varepsilon^{\rho i k}\varepsilon_{\rho j l}=\delta^i_j\delta^k_l-
\delta^i_l\delta^k_j~,$$
(where $\varepsilon$ is entirely antisymmetric unit tensor) we receive the interaction term
separating diquark channel \cite{cah} as
$$
{\vf \lambda}^i_j{\vf \lambda}^k_l=\left(4\alpha-
\frat{2}{3}\right)\delta^i_j\delta^k_l+
(2-4\alpha )\delta^i_l\delta^k_j+4\alpha~\varepsilon_{\rho ik}\varepsilon_{\rho lj}$$
where $\alpha$ is an arbitrary number.
Now making use the Fierz identity for color SU(3) matrices
$$\delta^i_j\delta^k_l=\frat{1}{3}\delta^i_l\delta^k_j
+\frat{1}{2}{\vf \lambda}^i_l{\vf \lambda}^k_j~,$$
we transform the interaction term to the following form:
$$
g~{\vf \lambda}^i_j{\vf \lambda}^k_l=g_s~\delta^i_l\delta^k_j+
g_{o}~{\vf \lambda}^i_l{\vf \lambda}^k_j+
g_d~\varepsilon^{\rho ik}\varepsilon_{\rho lj}~.
$$
$g_s=\frat{8}{3}\left(\frat{2}{3}-\alpha\right) g$, $g_o=2\left(\alpha-\frat{1}{6}\right) g$,
$g_d=4\alpha~g$,
which contains the singlet, octet and diquark terms.
Besides, the Fierz identities for the spin $\gamma$-matrices should be taken into account
$$\gamma^\mu_{\alpha\beta} \gamma^\mu_{\gamma\delta}= F^{{\mbox{\scriptsize{A}}}}
\gamma^{{\mbox{\scriptsize{A}}}}_{\alpha\delta} \gamma^{{\mbox{\scriptsize{A}}}}_{\gamma\beta}~,
$$
similarly for the diquark coupling we need
$$
\gamma^\mu_{\alpha\beta} \gamma^{{\scriptsize T}\mu}_{\delta\gamma}=-
F^{{\mbox{\scriptsize{A}}}} \gamma^{{\mbox{\scriptsize{A}}}}_{\alpha s} C_{s\gamma}\cdot
C_{\delta r}\gamma^{{\mbox{\scriptsize{A}}}}_{r\beta},$$
where the index ${\mbox{\scriptsize{A}}}$ refers to the channels: $1$, $\gamma_5$, $\gamma^\mu$,
$\gamma^\mu\gamma^5$, $F^{{\mbox{\scriptsize{A}}}}=1,-1,-1/2,-1/2$, correspondingly
(we take here that a permutation of quark fields does not result in changing the sign in
identity),
$C=\gamma^2\gamma^0$ is the charge conjugation matrix, $\gamma^{{\scriptsize T}}$ denotes
transposed matrix.
As a result the interaction term can be brought to the following form
\begin{eqnarray}
\label{8}
{\cal L}_{int}&=&g_s~F^{\mbox{\scriptsize{A}}}~
\bar q~\gamma^{\mbox{\scriptsize{A}}}~ q'\cdot \bar q'~\gamma^{\mbox{\scriptsize{A}}}~
q +\nonumber\\
&+&g_o~F^{\mbox{\scriptsize{A}}}~\bar q~\gamma^{\mbox{\scriptsize{A}}} {\vf \lambda}~ q'
\cdot \bar q'~\gamma^{\mbox{\scriptsize{A}}} {\vf \lambda}~q +\nonumber\\
&+&g_d~F^{\mbox{\scriptsize{A}}}~ \bar q ~\varepsilon^\rho \gamma^{\mbox{\scriptsize{A}}}
C~\bar q'
\cdot q'~\varepsilon_\rho C^{-1}\gamma^{\mbox{\scriptsize{A}}}~ q~,\nonumber
\end{eqnarray}
where the aggregated notations are used.
It is convenient to introduce the bispinors associated with antiquarks $\bar q_c$, $q_c$
while dealing with a color superconductivity
$$
q_c=C~\bar q^{{\mbox{\scriptsize T}}}~,~~\bar q_c=q^{{\mbox \scriptsize T}}C~.
$$
Remembering now the following identities
\begin{eqnarray}
\label{10}
&&\bar q q=\bar q_c q_c~,
~~\bar q \gamma^\mu q=-\bar q_c \gamma^\mu q_c~,\nonumber\\
&&\bar q \gamma^\mu \partial_\mu q=-\bar q_c \gamma^\mu
{\overleftarrow \partial_\mu} q_c=
\bar q_c \gamma^\mu
{\overrightarrow \partial_\mu} q_c~,\nonumber
\end{eqnarray}
the interaction term can be rewritten as
\begin{eqnarray}
\label{11}
&&{\cal L}_{int}=g_s~F^{\mbox{\scriptsize{A}}}~
\bar q~\gamma^{\mbox{\scriptsize{A}}}~ q'\cdot \bar q'~
\gamma^{\mbox{\scriptsize{A}}}~ q +\nonumber\\
&&+g_o F^{\mbox{\scriptsize{A}}}\bar q\gamma^{\mbox{\scriptsize{A}}} {\vf \lambda} q'
\cdot \bar q'~\gamma^{\mbox{\scriptsize{A}}}{\vf \lambda}~q %+\nonumber\\&&
-g_d F^{\mbox{\scriptsize{A}}}\bar q \varepsilon^\rho \gamma^{\mbox{\scriptsize{A}}}q'_c
\cdot \bar q'_c \varepsilon_\rho \gamma^{\mbox{\scriptsize{A}}}q.\nonumber
\end{eqnarray}
It is of importance to mention the change of sign in the last summand occurs due to the
identity valid for the charge conjugation matrix $C^{\mbox{\scriptsize T}}=-C$.
Now the singlet and octet components symmetric over $\bar q_c$, $q_c$ fields may be presented as
\begin{eqnarray}
\label{12}
&&\hspace{-0.3cm}{\cal L}_{int}=\frat{g_s}{4}F^{\mbox{\scriptsize{A}}}
[\bar q\gamma^{\mbox{\scriptsize{A}}} q'(\pm)^{\mbox{\scriptsize{A}}}
\bar q_c\gamma^{\mbox{\scriptsize{A}}} q'_c]\cdot
[\bar q'\gamma^{\mbox{\scriptsize{A}}}q (\pm)^{\mbox{\scriptsize{A}}}
\bar q'_c\gamma^{\mbox{\scriptsize{A}}}~q_c]
+\nonumber\\
&&\hspace{-0.3cm}+\frat{g_o}{4} F^{\mbox{\scriptsize{A}}}
[\bar q\gamma^{\mbox{\scriptsize{A}}}{\vf \lambda}q'(\pm)^{\mbox{\scriptsize{A}}}
\bar q_c\gamma^{\mbox{\scriptsize{A}}}{\vf \lambda}q'_c]
\cdot
[\bar q'\gamma^{\mbox{\scriptsize{A}}}{\vf \lambda}q (\pm)^{\mbox{\scriptsize{A}}}
\bar q'_c\gamma^{\mbox{\scriptsize{A}}}{\vf \lambda}q_c]+\nonumber\\
&&\hspace{-0.3cm}- g_d~ F^{\mbox{\scriptsize{A}}}~\bar q~ \varepsilon^\rho
 \gamma^{\mbox{\scriptsize{A}}}~ q'_c
\cdot \bar q'_c~ \varepsilon_\rho \gamma^{\mbox{\scriptsize{A}}}~  q~,\nonumber
\end{eqnarray}
sign $(\pm)^{\mbox{\scriptsize{A}}}$ is determined by the interaction channel
$\gamma^{\mbox{\scriptsize{A}}}$.
Using the quark doublets $\bar Q=(\bar q, \bar q_c)$, $Q=\left(\begin{array}{l}
q\\q_c\end{array}\right)$ the Lagrangian (\ref{7}) can be written as
\begin{eqnarray}
%8
\label{13}
{\cal L}&=&\frat12\bar Q~[(i\gamma_\mu \partial_\mu-m)\sigma_0+\hat \mu\sigma_3]~Q+
\nonumber\\[-.1cm]\\[-.1cm]
&+&\frat{g_s}{4}~ F^{\mbox{\scriptsize{A}}}~
\bar Q~\gamma^{\mbox{\scriptsize{A}}}\sigma_{\mbox{\scriptsize{A}}}~ Q' \cdot
\bar Q'~\gamma^{\mbox{\scriptsize{A}}}\sigma_{\mbox{\scriptsize{A}}}~Q+\nonumber\\
&+&\frat{g_s}{4}~ F^{\mbox{\scriptsize{A}}}~
\bar Q~\gamma^{{\mbox{\scriptsize{A}}}}{\vf \lambda}\sigma_{\mbox{\scriptsize{A}}}~Q'
\cdot \bar Q'~\gamma^{{\mbox{\scriptsize{A}}}}
{\vf \lambda}\sigma_{\mbox{\scriptsize{A}}}~Q+\nonumber\\
&-&g_d~F^{\mbox{\scriptsize{A}}}~\bar Q ~\varepsilon^\rho\gamma^{\mbox{\scriptsize{A}}}
\sigma_+~\bar Q'
\cdot \bar Q'~\varepsilon^\rho\gamma^{\mbox{\scriptsize{A}}} \sigma_-~ Q ~,\nonumber
\end{eqnarray}
where $\sigma_\pm=(\sigma_1\pm i\sigma_2)/2$. The matrices $\sigma$ re acting in the space of
variables $\bar Q$, $Q$, and $\sigma_0$ is a unit matrix, $\sigma_i$ are the Pauli matrices.
The notations of direct products of matrices were omitted here and it was meant that either the
matrix $\sigma_0$ or $\sigma_3$ appear in this expression in dependence on the channel $A$, but
for simplicity we indicate by this notation that a concrete form of the matrix
$\sigma_{\mbox{\scriptsize{A}}}$ should be specified.

The mean field approximation assumes an identification of non-trivial vacuum expectation values
and formulation of respective self-consistency conditions. Now we consider several particular
examples, first of all, for the normal conditions ($T=0$, $\mu=0$) without resorting the
approximations related to separating out the dominant interaction (as it is usually done),
and try to take into account the contributions of all channels exactly. It becomes clear later
this task is very hard to be completed accurately but it is possible to cover almost an entire
spectrum of options by analysing the channels.

\subsection*{Nontrivial average in scalar channel}
For consistency, we will first reproduce the already
known result for the quark dynamical mass. Let the
nontrivial vacuum expectation value is generated in a
scalar channel
$$
{\cal L}_{int}\simeq 2 g_s  ~\bar q q' ~\langle \mbox{Tr}\bar q' q\rangle=
-2 g_s~\bar q q'~ \mbox{Tr}\{i S \}~.
$$
Then using the Green function $S=-i\langle q \bar q'\rangle$,  $S^{-1}=\hat p-M_q$,
for the KKB model one can obtain
$$
M= 2 g_s ~\mbox{Tr}\left\{\frac{1}{\hat p-M_q}\right\}=
2 G_s~ \frac{M_q}{E_p}~,
$$
$E_p=({\vf p}^2+M_q^2)^{1/2}$, $G_s=2 N_c g_s$. Here,
taking the trace means also integration in $\int d p_0 i/(2\pi)$.
The expression deduced coincides with Eq. (\ref{6}).

We need also the energy density of the quark ensemble $E= 2 N_c w$,
which can be expressed through specific energy attributed to a one quark, see \cite{MZ2},
in form
\begin{eqnarray}
%9
\label{16}
w&=&\int d \widetilde{\vf p}~p_0-\int d \widetilde{\vf p}~(1-n-\bar n)~P_0+
\nonumber\\[-.2cm]
\\ [-.25cm]
&+&\frac{1}{4G_s}~\int d \widetilde{\vf p}d \widetilde{\vf q}~~
F({\vf p}+{\vf q})~\widetilde M({\vf p})
\widetilde M({\vf q})~.\nonumber
\end{eqnarray}
In this expression a density of induced quark mass is used
\begin{equation}
%10
\label{17}
M_q({\vf p})=m+M({\vf p})=m+\int d \widetilde{\vf q}~
F({\vf p}+{\vf q})~\widetilde M({\vf q})~.
\end{equation}
The density of mean quark ensemble energy  (\ref{16}) is nothing more than the energy functional
of the Landau's Fermi-liquid theory \cite{MZ3} (see also \cite{TM}), variation of which
over density of induced quark mass gives equation for dynamical quark mass (\ref{4}).

\subsection*{Scalar and pseudoscalar channels}
Let us suppose non-trivial contributions occur in scalar and pseudoscalar channels:
\begin{eqnarray}
\label{p10}
&&{\cal L}_{int}\simeq 2 g_s~ \bar q q' ~\langle \mbox{Tr}\bar q' q\rangle
+2g_s~\bar q i\gamma^5 q' ~\langle \mbox{Tr}\bar q' i\gamma^5 q\rangle
=\nonumber\\
&&=-2 g_s~ \bar q q'~ \mbox{Tr}\{i S \}-
2 g_s~ \bar q i\gamma^5 q'~ \mbox{Tr}\{i\gamma^5 i S \}~.\nonumber
\end{eqnarray}
Now the Green function is determined by the relation
$$S^{-1}=\hat p-M_q-i\gamma^5 C_q~.$$
Then the selfconsistency relations take the form
\begin{eqnarray}
%11
\label{p11}
&&M=2 G_s ~\mbox{Tr}\left\{\frac{1}{\hat p-M_q-i\gamma^5 C_q}\right\}=
2 G_s~ \frac{M_q}{E_p}~,\nonumber\\[-.1cm]\\[-.1cm]
&&C=2 G_s~\mbox{Tr}\left\{\frac{i\gamma^5}{\hat p-M_q-i\gamma^5 C_q}\right\}=
2 G_s~\frac{C_q}{E_p}~,\nonumber
\end{eqnarray}
where $C_q=c+C$, $E_p=({\vf p}^2+M_q^2+C_q^2)^{1/2}$.
Here $c$ is a bare ("current"{\phantom{.}})
"mass"{\phantom{.}} in pseudoscalar channel. At $c\equiv 0$, $C=0$, because in general case
it is impossible to satisfy both conditions (\ref{p11}) for dynamical masses $M$, $C$.
Let us stress that at $c\equiv 0$, $m\equiv 0$ it turns out impossible to determine
the phase of dynamical mass. However, if the Bogolyubov procedure of quasi-averages is applied,
for example, in the form $m\to 0$, $c\equiv 0$, the phase is precisely restored. If the bare
pseudoscalar mass $c$ is taken to be of quark current mass $m$ order, then the induced mass $C$
corresponds to the standard values of the quark induced mass $M$ order.

\subsection*{Scalar and Isotriplet of the Octet Channel for SU(3) group}
Let us select the following averages in this configuration:
\begin{eqnarray}
\label{p12}
&&{\cal L}_{int}\simeq 2g_s~\bar q q' ~\langle \mbox{Tr}\bar q' q\rangle
+2 g_o\bar q \frac{{\vf \lambda}}{2} q'
~\langle \mbox{Tr}\bar q' \frac{{\vf \lambda}}{2} q\rangle
=\nonumber\\
&&=-2 g_s~\bar q q'~ \mbox{Tr}\{i S \}-
2 g_o~ \bar q {\vf \lambda} q'~
\mbox{Tr}\{ {\vf \lambda} i S \}~.\nonumber
\end{eqnarray}
The Green function is found from the following relation
$$S^{-1}=\hat p-M_q- {\vf \lambda} {\vf O}_q~.$$
For the sake of simplicity we suppose that only the following isotriplet components
$\lambda_i$, $i=1,2,3$ of the vector ${\vf O}_q$ are involved. Then the Green function can be
presented in the form
\begin{eqnarray}
\label{p13}
&&S=A+{\vf B} {\vf \lambda}~,\nonumber\\
&&A=\frac12 \left\{\frac{1}{\hat p-M_q+O_q}+\frac{1}{\hat p+M_q+O_q}\right\}~,\nonumber\\
&&{\vf B}=-\frac12\frac{{\vf O}_q}{O_q}
 \left\{\frac{1}{\hat p-M_q+O_q}-\frac{1}{\hat p+M_q+O_q}\right\}~,\nonumber
\end{eqnarray}
$O_q=|{\vf O}_q|$ and it allows us to find the self-consistency conditions as
\begin{eqnarray}
\label{p14}
&&M=2 g_s ~\mbox{Tr}\frac12\left\{A \right\}~,\nonumber\\
&&O_i=2g_o ~\mbox{Tr}\left\{\lambda_i ({\vf B}{\vf \lambda})\right\}~.\nonumber
\end{eqnarray}
Finally we obtain
\begin{eqnarray}
\label{p15}
&&M=2 G_s ~\frac12\left\{\frac{M^+}{E^+_{\vf p}}+
\frac{M^-}{E^-_{\vf p}}\right\}~,\nonumber\\
&&O=2 G_o\left\{\frac{M^+}{E^+_{\vf p}}-
\frac{M^-}{E^-_{\vf p}}\right\}~,\nonumber
\end{eqnarray}
where $G_o=N_c g_o$, $M_\pm=M_q\pm O_q$, ${\vf O}_q={\vf o}+{\vf O}$,
($O$ is the projection of vector {\vf O} upon some selected direction),
$E^\pm_p=({\vf p}^2+M_\pm^2)^{1/2}$.
If the "current"{\phantom{.}} value of vector ${\vf o}$ equals to zero, a trivial
solution with zero dynamical vector ${\vf O}$ does exist and then $M_q^+=M_q^-$.
An insignificant admixture of octet channel, $|{\vf o}| \ll m$, leads to the small contributions
for the quark dynamical mass.

\subsection*{Diquark Condensation (Color Superconductivity), $\mu \neq 0$, $T=0$}
We restrict ourselves here with considering the color superconductivity in pseudoscalar channel
$\gamma_5$, where, as we demonstrate below, there exist an interesting solution with a real gap.
The self-consistency relations in this case take the form
\begin{eqnarray}
%n1
\label{n1}
&&M=-2g_s~ \mbox{Tr}\langle \bar q q' \rangle~,~~\nonumber\\
&&\Delta^\rho=-2 g_d~\mbox{Tr}\langle \bar q'_c \varepsilon^\rho \gamma_5 q \rangle~,~~
\stackrel{*}{\Delta}^\rho=2 g_d~\mbox{Tr}\langle \bar q \varepsilon^\rho \gamma_5 q'_c \rangle~.\nonumber
\end{eqnarray}
In order to simplify an analysis we select the third component $\rho=3$ as nontrivial for color
superconductivity. The matrix of inverse Green function has the form

\begin{figure}%[!tbh]
\includegraphics[width=0.3\textwidth]{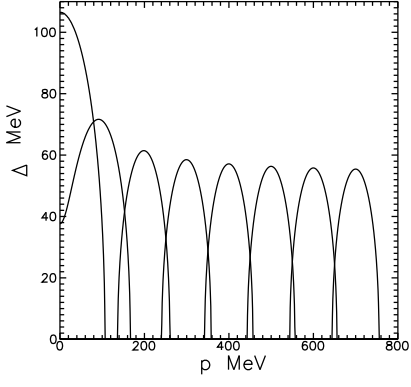}
\caption{Gap as a function of the momentum of the quark for the KKB model. Current quark mass is $m=5$ MeV.
The leftmost curve is obtained for normal conditions.  The following right ones are received
for the growing chemical potential with the step $100$ MeV.
}
\label{f3}
\end{figure}

\vspace{0.25cm}
\begin{center} \label{rot}
\parbox[b] {3.in} {$% {3.6in}
{\cal S}^{-1} = \left\| \begin{array} {cc}
                       (\hat p_+-M_q)~\delta_{ij}&\Delta^\rho \varepsilon^{\rho ij}\gamma_5\\
          \stackrel{*}{\Delta}^\rho \varepsilon^{\rho ij} \gamma_5
                                                 &(\hat p_--M_q)~\delta_{ij}
\end{array}
\right\|. $ }
\end{center}
\vspace{0.25cm}
\noindent
Such a form of the matrix is a direct consequence of the spinor $\bar q_c$, $q_c$ forms. The
Green's function itself is obtained from the Frobenius formula for block matrices\\
${\cal S}^{-1}=\left|\left|\begin{array}{cc}
A&I\\
J&B
\end{array}\right|\right|$~,~~${\cal S}=\left|\left|\begin{array}{cc}
K^{-1}&-A^{-1}I L^{-1}\\
-L^{-1}J A^{-1}&L^{-1}
\end{array}\right|\right|$~.\\
There exist two diagonal in color components $A=(\hat p_+-M_q)~\delta_{ij}$,
$B=(\hat p_--M_q)~\delta_{ij}$, and two components including antisymmetric in indices tensor
 $I=\Delta^3 \varepsilon^{3ij} \gamma_5$,
$J=\stackrel{*}{\Delta^3} \varepsilon^{3ij} \gamma_5$. In order to proceed we need to consider
the blocks of the Green function construction
\begin{eqnarray}
K&=&A-I B^{-1}J=K_\sigma~\Sigma_{ij}+K_d~D_{ij}~,\nonumber\\
&&\hspace{-0.75cm}K_\sigma=\hat p_+-M_q,~K_d=\left [\hat p_+-M_q -\Delta^2
\frat{\hat p_--M_q}{(p_0-\mu)^2-E_p^2} \right].\nonumber
\end{eqnarray}
To get this expression we use the fact that product of two antisymmetric tensors generates
(doublet) projector upon the  $1$ and $2$ components of the color space
$$D_{ij}=\delta_{1i}\delta_{j1}+\delta_{2i}\delta_{j2}=-\varepsilon^{3ik}\varepsilon^{3kj}~.$$
Then, we have introduced an additional singlet component
$$\Sigma_{ij}=\delta_{3i}\delta_{j3}~,~~\delta_{ij}=\Sigma_{ij}+D_{ij}~,$$
and decomposed the matrix block $K$ into two parts--the singlet $K_\sigma$ and the
doublet $K_d$. In principle, a similar expansions should be done for the components $A$ and $B$
but we omit this complication hoping it does not lead to confusion.
Here $\Delta^2=\Delta^3 \stackrel{*}{\Delta^3}$ is a convenient abbreviation of notation for
the positive quantity. Similarly
$$L=B-J A^{-1} I=L_\sigma~\Sigma_{ij}+L_d~D_{ij}~,$$
where the tensor components are
$$L_\sigma=\hat p_--M_q~,~~L_d=\hat p_--M_q-\Delta^2\frat{\hat p_+-M_q}{(p_0+\mu)^2-E_p^2}~.$$
Having found the inverse components $K_\sigma^{-1}$, $L_\sigma^{-1}$, in particular,
$$K_d^{-1}=[(p_-^2-M_q^2)(\hat p_++M_q)-\Delta^2(\hat p_-+M_q)]/Z_{\mbox{\scriptsize{T}}}~,$$
$$L_d^{-1}=[(p_+^2-M_q^2)(\hat p_-+M_q)-\Delta^2(\hat p_++M_q)]/Z_{\mbox{\scriptsize{T}}}~,$$
$$Z_d=[p_0^2-(E+\mu)^2-\Delta^2][p_0^2-(E-\mu)^2-\Delta^2]~,$$
where $p_\pm^2=\hat p_\pm \hat p_\pm$,
we are able to obtain the following expressions for the building elements of the Green function
$$-A^{-1}I L^{-1}=\gamma_5 \Delta^3~[L_d (\hat p_++M_q)]^{-1}~\varepsilon^{3ij}~,$$
$$-L^{-1}J A^{-1}=\gamma_5 \stackrel{*}{\Delta^3}~[(\hat p_++M_q)L_d]^{-1}~\varepsilon^{3ij}~.$$
Utilizing these results we can expand the Green function in the components of the matrix
$\sigma_i$ as
\begin{eqnarray}
%12
\label{p16}
&&\hspace{-0.5cm}{\cal S}=\sum_{i=0,\cdots 3} S^i~\sigma_i~,\\
&&\hspace{-0.5cm}S^0=S^0_\sigma~\Sigma_{ij}+S^0_d~D_{ij}~,\nonumber\\
&&\hspace{-0.5cm}S^0_\sigma=[(p_0^2+\mu^2-E^2_p)(\hat p+M_q)
-2\mu p_0~\hat \mu]/Z_\sigma~,\nonumber\\
&&\hspace{-0.5cm}S^0_d=[(p_0^2+\mu^2-E^2_p-\Delta^2)(\hat p+M_q)
-2\mu p_0~\hat \mu]/Z_d~,\nonumber\\
&&\hspace{-0.5cm}S^1+S^2=\gamma^5\Delta^3\varepsilon^{3ij}(p_0^2-\mu^2-E^2_p-\Delta_q^2+
\hat\mu\hat p-\hat p\hat \mu)/Z_d,\nonumber\\
&&\hspace{-0.5cm}S^1-S^2=\gamma^5\stackrel{*}{\Delta^3}\varepsilon^{3ij}(p_0^2-\mu^2-E^2_p-
\Delta_q^2-\hat\mu\hat p+\hat p\hat \mu)/Z_d,\nonumber\\
&&\hspace{-0.5cm}S^3=S^3_\sigma~\Sigma_{ij}+S^3_d~D_{ij}~,\nonumber\\
&&\hspace{-0.5cm}S^3_\sigma=[(p_0^2+\mu^2-E^2_p)~
\hat \mu-2\mu p_0~(\hat p+M_q)]/Z_\sigma~,\nonumber\\
&&\hspace{-0.5cm}S^3_d=[(p_0^2+\mu^2-E^2_p+\Delta_q^2)~
\hat \mu-2\mu p_0~(\hat p+M_q)]/Z_d~,\nonumber
\end{eqnarray}
and
$$Z_\sigma=[(p_0+\mu)^2-E_p^2][(p_0-\mu)^2-E_p^2]~.$$
calculating the traces, which include the integration over the component $p_0$,
and allocating channels in the self-consistency equation (\ref{n1}),
finally we obtain the following decomposed system of equations
\begin{eqnarray}
%13
\label{p21}
&&M=4~g_s~\frat{M_q}{E_p}~,\nonumber\\
&&M=4~g_s~\frat{M_q}{E_p}~\left[\frac{E_p+\mu}{P_+}+\frac{E_p-\mu}{P_-}\right]~,\\
&&1=4~g_d~  \left[\frac{1}{P_+}+\frac{1}{P_-}\right]~,\nonumber
\end{eqnarray}
where $P_\pm=[(E_p\pm\mu)^2+\Delta^2]^{1/2}$. The first line describes the third (singlet)
component in the color space. Comparison it with the equation (\ref{6}) we conclude that in
general case the diquark channel has no any impact on the dynamical mass, only the coupling
constant is getting three times smaller value. The second and third lines describe the color
components of doublets $1$ and $2$. It is reasonable to mention here that we have not introduced
the bare condensate $\Delta_0$ and it is why the third equation does not contain a gap
explicitly. It is also convenient to introduce the following designations of the coupling
constants $G_s=2~N_c g_s$, $G_d=2\cdot 2 g_d$. One curios fact of this machinary is that if
the scalar channel is used for color superconductivity instead of the pseudoscalar one the system
of equations (\ref{p21}) becomes inconsistent (controversial) because in the lowest line a
negative unity appears instead of positive unity. We do not introduce the additional separate
notations for the masses of singlet and doublet channels in order not to overload the formulae.
A thorough analysis of this equation system shows the system has no the consistent real
solutions, and they appear only when either the induced mass or the gap are getting zero values
separately. The easiest way to observe this fact is to analyse the particular case when, for
example, $\mu=0$, i.e. in practice such a complex equation system is unnecessary, and we return
to the equation (\ref{6}) at $\Delta=0$, and have the third line of the system (\ref{p21})
valid when the induced quark mass is zero ($M=0$). In order to have a reasonable estimate of the
gap characteristic values we consider the particular situation of normal conditions with the
parameter $\alpha=1/6$. The octet channel contribution disappears for such a configuration.
Setting up the coupling constant $g$ such a value that the dynamical quark mass at zero momentum
for normal conditions ($\mu=0$) at zero temperature in the chiral condensate phase equals to
$M_q(0)\approx 345$ MeV we have from the third line of the system (\ref{p21}) for the gap energy
$P=2G_d \approx 114$ MeV. In the chiral limit $m=0$, $M_q(0) \approx 340$ MeV then
$$\Delta=(4G_d^2-p^2)^{1/2}~.$$
In Fig. \ref{f3} the solutions for the gap equation are displayed when the current quark mass is
$m=5$ MeV. The leftmost curve is obtained for normal conditions. The following right ones
were received for the growing chemical potential with the step $100$ MeV.
%%%%%%%%%%%%%%%%%%%%%%%%%%%%%%%%%%%%%%%%%%%%%%%%%
%%%%%%%%%%%%%%%%%%%%%%%%%%%%%%%%%%%%%%%%%
It is clear that, although the color superconductivity can be developed for normal conditions,
the  gap value is essentially smaller (approximately three times less) than the chiral condensate.
However, at the chemical potential growing the dynamical quark mass in the phase of non-zero chiral
condensate falls quickly down and one could expect the color superconductivity phase more
advantageous at large chemical potentials. In order to clarify this point we need to compare the
energies and baryon charges of both phases. We are doing that here but in the chiral limit only.
Then the quark energy in the phase of non-zero chiral quark condensate is degenerate as it follows
from equation (\ref{6}), $E_p=2G_s$. Therefore, to deal with the quark ensemble of changing
a baryon/quark density we need to introduce the Fermi momentum $P_{\mbox{\scriptsize{F}}}$ that
characterises the process of filling up the Fermi sphere by quark quasiparticles (the details
can be found in \cite{MZ2}). Having determined the energy density and baryon charge density of
an ensemble we may obtain a inter-relation between the Fermi momentum and the chemical potential.
It is worth to remember here the dynamical quark mass has the zero value for momenta smaller than
the Fermi momentum of $M_q=0$, $p<P_{\mbox{\scriptsize{F}}}$ in the chiral limit. Beyond the
Fermi momentum the dynamical quark mass is defined as $M_q=(4G_s^2-p^2)^{1/2}$,
$P_{\mbox{\scriptsize{F}}}<p<2G_s$. At large momenta $2G_s<p$  the dynamical quark mass has the
zero value. The ensemble energy density is defined as (we have omitted an unimportant
normalization constant, i.e. the first term in Eq.(\ref{16}))
\begin{eqnarray}
%14
\label{n2}
&&H_{cc}=\int\limits_{P_{\mbox{\scriptsize{F}}}}^{2G_s}\!\!d \widetilde{\vf p}~
\left( -2N_c~P_0+\frat{M^2}{4g_s}\right)=\nonumber\\[-.1cm]\\ [-.1cm]
&&=-\frat{N_c}{3\pi^2}~2G_s~[(2G_s)^3-P_{\mbox{\scriptsize{F}}}^3]+\nonumber\\
&&+\frat{1}{8\pi^2 g_s}\left(\frat{2(2G_s)^5}{15}-\frat{(2G_s)^2P_{\mbox{\scriptsize{F}}}^3}{3}
+\frat{P_{\mbox{\scriptsize{F}}}^5}{5}\right)
,~P_{\mbox{\scriptsize{F}}}<2G_s\nonumber\\
&&\hspace{-0.5cm}H_{cc}=\frat{N_c}{4\pi^2}~[P_{\mbox{\scriptsize{F}}}^4-(2G_s)^4]~,
~~P_{\mbox{\scriptsize{F}}}>2G_s.\nonumber
\end{eqnarray}
The baryon number density is defined as
\begin{equation}
%15
\label{n3}
Q_{cc}=2N_c\int\limits^{P_{\mbox{\scriptsize{F}}}}
d \widetilde{\vf p}=\frat{N_c}{3\pi^2}~P_{\mbox{\scriptsize{F}}}^3~.
\end{equation}
Now we are able to define the energy density and baryon number density of the superconducting
state $H_{sc}$. By definition, the energy is given by
\begin{eqnarray}
%16
\label{n4}
\hspace{-0.25cm}H_{sc}\!\!&=\!\!&\!\!\frat12\!\!
\int d \widetilde{\vf p}\!\!\int\frat {dp_0 (-i)}{2\pi}
\mbox{Tr} \{ \gamma^0 p^0\otimes \sigma_0\otimes [\Sigma+D]{\cal S}
e^{-ip\varepsilon}\}+\nonumber\\
&+&\int d \widetilde{\vf p}~\frat{\Delta^2}{4g_d}.
\end{eqnarray}
Here, the direct products of the free Hamiltonian part pick out a spinor structure, we are
interested in, $\gamma_0$, unit matrix $\sigma_0$ acting in the quark doublet space and,
finally, an identity matrix of color space decomposed in the singlet $S$ and doublet $D$
components. The second term describes the interaction energy and may be obtained in a way
similar to the interaction contribution in Eq. (\ref{16}) (similar Eq. (\ref{n2})) by
calculating the average energy over the state which we are interested in. Making use the
expansion of matrix ${\cal S}$ (\ref{p16}) and calculating the corresponding integrals we
obtain
\begin{equation}
%17
\label{n5}
H_{sc}=H_\sigma+H_d~,
\end{equation}
where $H_\sigma$ is the contribution of singlet component (3) in color space and $H_d$ is
the contribution of doublet $1$, $2$. The color component 3 being pure as to the color
superconducting condensate can be easily calculated as
\begin{equation}
%18
\label{n6}
H_\sigma=2\int\limits^{P_{\mbox{\scriptsize{F}}}}\!\!d \widetilde{\vf p}~ p=
\frat{1}{4\pi^2}~P_{\mbox{\scriptsize{F}}}^4~.
\end{equation}
It looks interesting to note this contribution is in factor $N_c$ weaker than the
contribution of $H_{cc}$ because of the color singlet lost. Moreover, as it was mentioned,
in the color superconducting phase the dynamical quark mass equals precisely to zero
because the self-consistent solutions of the system (\ref{p21}) are absent, i.e. the quark
energy contribution is $p$. The contribution of the quark doublet looks like
\begin{equation}
%19
\label{n7}
H_d=-2\cdot 2\int d \widetilde{\vf p} ~\frat12~(P_++P_-)+
\int d \widetilde{\vf p}~\frat{\Delta^2}{4g_d}~.
\end{equation}
Here in the free part (the first term) one factor 2 corresponds to the contribution of two
color components and another factor 2 corresponds to the contribution of the two spin
components. It is also interesting to note that these expressions are the energy functionals
of the Landau theory of Fermi liquid, and and the equation system (\ref{p21}) can be
calculated by taking their variations over dynamical mass and gap.
Similarly, we have for the baryon charge
\begin{eqnarray}
%20
\label{n8}
&&Q_{sc}=\frat12 \int d \widetilde{\vf p}\int\frat {dp_0 (-i)}{2\pi}
\mbox{Tr} \{ \gamma^0 \otimes \sigma_3\otimes [\Sigma+D]{\cal S}
e^{-ip\varepsilon}\}.\nonumber\\
&&Q_{sc}=Q_\sigma+Q_d~,\\
&&Q_\sigma=\frat{1}{3\pi^2}~P_{\mbox{\scriptsize{F}}}^3~,\nonumber\\
&&Q_d=2\cdot 2\int d \widetilde{\vf p}
~\frat12~\left(\frat{E_p+\mu}{P_+}-\frat{E_p-\mu}{P_-}\right)~.\nonumber
\end{eqnarray}
Comparing the bottom line to Eq. (\ref{n7}) we find the following identity
$$Q_d=-\frat{\partial H_d}{\partial \mu}~,$$
because  only first summand of the type of free energy shows an obvious dependence on the
chemical potential in Eq. (\ref{n7}).
%%%%%%%%%%%%%%%%%%%%%%%%%%%%%%%%%%%%%%%%%%%%%%%%%
\begin{figure}%[!tbh]
\includegraphics[width=0.3\textwidth]{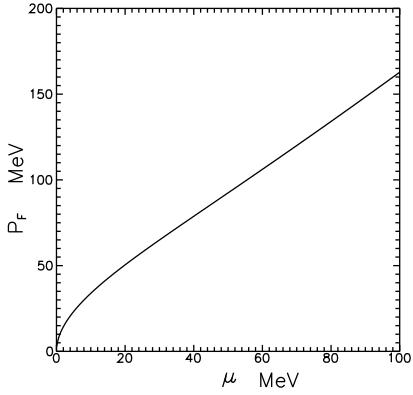}
\caption{The Fermi momentum as a function of the chemical potential for the third color
component on in the phase of color superconductor, see  text.
}
\label{f4}
\end{figure}
%%%%%%%%%%%%%%%%%%%%%%%%%%%%%%%%%%%%%%%%%
Now we find the inter-relation between the Fermi momentum and chemical potential
$P_{\mbox{\scriptsize{F}}}(\mu)$.
By definition, a chemical potential is an energy necessary to add (remove) a quasiparticle
into a system
\begin{equation}
%21
\label{n10}
\mu=\frat{d H}{d Q}=\frat{\frat{\partial H}{\partial P_{\mbox{\scriptsize{F}}}}
d P_{\mbox{\scriptsize{F}}}+\frat{\partial H}{\partial \mu}~ d\mu}
{\frat{\partial Q}{\partial P_{\mbox{\scriptsize{F}}}}
d P_{\mbox{\scriptsize{F}}}+\frat{\partial Q}{\partial \mu}~ d\mu}~.
\end{equation}
It allows us to obtain the differential equation for the function
$P_{\mbox{\scriptsize{F}}}(\mu)$
\begin{equation}
%22
\label{n11}
\frat{d P_{\mbox{\scriptsize{F}}}}{d \mu}
=-\frat{\mu\frat{\partial Q}{\partial P_{\mbox{\scriptsize{F}}}}
-\frat{\partial H}{\partial P_{\mbox{\scriptsize{F}}}}}
{\mu\frat{\partial Q}{\partial \mu}
-\frat{\partial H}{\partial \mu}}~.
\end{equation}
In the case of interest, the obvious dependence on the Fermi momentum is available in
the contributions $H_\sigma$ and $Q_\sigma$. The free part of energy $H_d$ and baryon
charge $Q_d$ are obviously depend on the chemical potential
$$\frat{\partial Q}{\partial P_{\mbox{\scriptsize{F}}}}=
\frat{\partial Q_\sigma}{\partial P_{\mbox{\scriptsize{F}}}}=
\frat{P_{\mbox{\scriptsize{F}}}^2}{\pi^2}~,~~
\frat{\partial H}{\partial P_{\mbox{\scriptsize{F}}}}=
\frat{\partial H_\sigma}{\partial P_{\mbox{\scriptsize{F}}}}=
\frat{P_{\mbox{\scriptsize{F}}}^3}{\pi^2}~.$$
$$\frat{\partial Q}{\partial \mu}=\frat{\partial Q_d}{\partial \mu}=
2\cdot 2\int d \widetilde{\vf p} ~\frat12~\Delta^2~\left(\frat{1}{P_+^3}+\frat{1}{P_-^3}\right)~.$$
Using these relations we obtain
\begin{equation}
%23
\label{n12}
\frat{P_{\mbox{\scriptsize{F}}}^2~(P_{\mbox{\scriptsize{F}}}-\mu)}{\pi^2}~
\frat{d P_{\mbox{\scriptsize{F}}}}{d\mu}=
\mu\frat{\partial Q_d}{\partial \mu}+Q_d=
\frat{\partial \mu~ Q_d}{\partial \mu}~.
\end{equation}
which results in the definition of the Fermi momentum as
\begin{equation}
%24
\label{n13}
P_{\mbox{\scriptsize{F}}}^4-\mu~P_{\mbox{\scriptsize{F}}}^3-3\pi^2~\mu Q_d=0~.
\end{equation}
We can also consider Eq. (\ref{n12}) as a differential equation with the natural
initial condition $P_{\mbox{\scriptsize{F}}}(0)=0$. In particular, at $\mu\to 0$ we can obtain
$Q(0)=0$, $\partial Q/\partial \mu=4 (2G_d)^2/(15\pi^2)$. Then for the Fermi momentum we have
approximately
$$P_{\mbox{\scriptsize{F}}}^4\approx \frat{4}{5}~(2G_d)^2~\mu^2~.$$
At large $\mu$ the Fermi momentum is linearly increasing, see Fig. \ref{f4}.
%%%%%%%%%%%%%%%%%%%%%%%%%%%%%%%%%%%%%%%%%%%%%%%%%
\begin{figure}%[!tbh]
\includegraphics[width=0.3\textwidth]{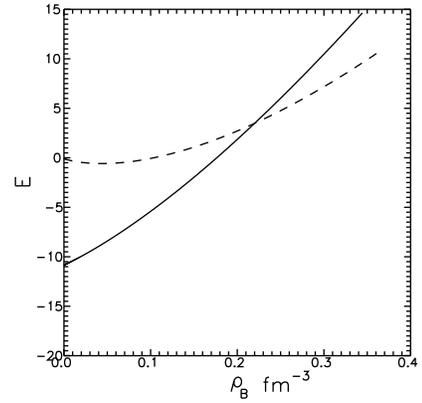}
\caption{The energy of the chiral phase (solid line) and the phase of color superconductivity
(dashed curve) as a functions of baryon number density.
}
\label{f5}
\end{figure}
%%%%%%%%%%%%%%%%%%%%%%%%%%%%%%%%%%%%%%%%%

In order to find the density of the baryon charge $Q_d$ (and its derivative
$\partial Q/\partial \mu$) we need to find out the boundary of momentum integration area with
nontrivial superconducting color condensate. Taking $\Delta$ in the third equation of
system (\ref{p21}) the zero value limit we obtain
$$p_{min}=0~,~\mu<2G_s~,~p_{min}=\mu \left(1-\frat{2G_s}{\mu}\right)^{1/2}~,~\mu>2G_s,$$
$$p_{max}=2G_s+\left(4G_s^2+\mu^2\right)^{1/2}~.$$
Fig. \ref{f5} demonstrates the energy (in an arbitrary unit) of the phase of non-zero chiral
condensate (solid line) and color superconducting phase (dashed line) as a function of baryon
number density which by definition is in factor three smaller than the density of quark baryon
charge $Q_{\mbox{\scriptsize{B}}}=Q/3$. It is visible that at low densities a formation of
chiral condensate is advantageous. At densities slightly higher than the density of normal
nuclear matter $\sim 0.217$/fm$^3$ ($P^c_{\mbox{\scriptsize{F}}}\approx 364$ MeV,
$\mu^c\approx 330$ MeV and for comparison the dynamical quark mass is $M(0)=2 G_s=340$ MeV)
the color superconductivity state becomes profitable. It is interesting to emphasize an
important role of the singlet component. The dashed line is quickly going down without its
contribution and the color superconductivity phase would be dominating already at unreasonably
low baryon number density.

We obtained the estimate of the chemical potential (the density of the quark ensemble) in the
chiral limit and expect that reaching this value a system could undergo the phase transition
into a color superconducting phase. However, beyond the chiral limit the situation remains
somewhat unclear. The energy of quark ensemble in a state with non-zero value of chiral
condensate is given by the following expression
similar to (\ref{n2})
\begin{equation}
%25
\label{n25}
H_{cc}=\int \widetilde{\vf p}~ 2N_c~p_0+\int\limits_{P_{\mbox{\scriptsize{F}}}}\!\!d \widetilde{\vf p}~
\left( -2N_c~P_0+\frat{M^2}{4g_s}\right)~,
\end{equation}
where $p_0=(p^2+m^2)^{1/2}$, $P_0=(p^2+M_q^2)^{1/2}$. The first term appears here because of a
normalization reasons in order to keep zero energy of ensemble at switching of the interaction.
Now turning to Eq. (\ref{6}) we conclude that the asymptotic behavior of induced quark mass and
its energy at large momenta $p\to \infty$ are the following
$$M\to 2 G_s \frat{m}{p_0}~,~~P_0\to p_0+\frat{Mm}{p_0}~.$$
Substituting the asymptotic expression in the (\ref{n25}), and bearing in mind the definition of
coupling constant $G_s=2N_c g_s$, we conclude that the ensemble energy diverges linearly,
and is going to the negative infinity but in the chiral limit the ensemble energy is finite.
Discontinuity of the functional of average energy as a function of current mass has been found
in \cite{MZ} where it was mentioned that any formal conclusion about the color superconductor
phase based on comparing the energy of two phases is unreliable. But it is worth of noting that
the infinite energy comes from integration with asymptotically low value of chiral condensate, and
then it looks quite possible to consider a mixed state at large momentum but an analysis of its
formation dynamics is beyond the scope of this our study.

%%%%%%%%%%%%%%%%%%%%%%%%%%%%%%%%%%%%%%%%%%%%%%%%%
\begin{figure}%[!tbh]
\includegraphics[width=0.3\textwidth]{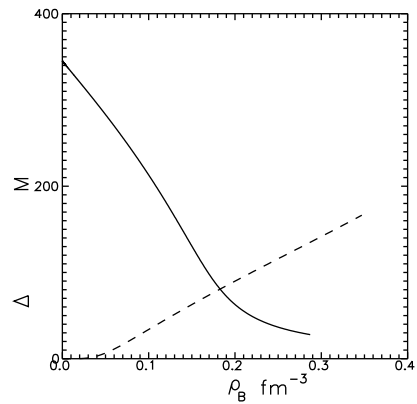}
\caption{Dynamical quark mass (solid line) and the gap (dashed line)
 as a functions of the baryon density in the NJL model.
}
\label{f6}
\end{figure}
%%%%%%%%%%%%%%%%%%%%%%%%%%%%%%%%%%%%%%%%%
In this article we consider a special choice of the parameter $\alpha$ responsible for the separation
of singlet, octet and diquark channels of interaction in order to neutralize the octet channel. Obviously,
it would be interesting to find out a way of fixing it grounded on the argument of the energy gain.

\subsubsection*{NJL model}
Now to have a deeper insight we present results for the model with a gluon correlator behaving as
a delta-function in coordinate space. The corresponding equation system the dynamical mass and gap looks like
\begin{eqnarray}
%26
\label{p26}
&&M=4~g_s~\int\limits^\Lambda d \widetilde{\vf p}~ \frat{M_q}{E_p}~,\nonumber\\
&&M=4~g_s~\int\limits^\Lambda d \widetilde{\vf p}~
\frat{M_q}{E_p}~\left[\frac{E_p+\mu}{P_+}+\frac{E_p-\mu}{P_-}\right]~,\\
&&\Delta=4~g_d~\Delta_q~
\int\limits^\Lambda d \widetilde{\vf p}~\left[\frac{1}{P_+}+\frac{1}{P_-}\right]~,\nonumber
\end{eqnarray}
and we suppose the quark mass and the superconductor gap do not depend on momentum. As in previous section
solutions are searched either in the form of the chiral condensate $M\neq 0$, $\Delta=0$, or in the form of
color superconductor $M=0$, $\Delta \neq 0$, and $\Delta_q=\Delta+\Delta_o$ where $\Delta_o$ describes the
current bare condensate. The equation for the dynamical mass in the phase of non-zero chiral condensate has
the following form
\begin{equation}
%27
\label{p27}
\hspace{-0.15cm}M=\frat{G_s}{2\pi^2}~M_q~\left(\Lambda E_\Lambda-P_{\mbox{\scriptsize{F}}}
E_{\mbox{\scriptsize{F}}}-M_q^2\ln \frat{\Lambda+E_\Lambda}{P_{\mbox{\scriptsize{F}}}+
E_{\mbox{\scriptsize{F}}}}\right).
\end{equation}
It is essential to keep in mind that the coupling constant $G_s$, and some others, which are present in
this consideration have a different dimension comparing to the coupling constants handled in previous
section. The constant $G_s$ is chosen in such a way to have the induced quark mass equal to $M=340$ MeV.
The parameter $\alpha=1/6$ fixes the inter-relation between the constant magnitudes in different channels,
similarly to what we have for the KKB model. The quark energy at given momentum denotes with a
corresponding index, for example $E_\Lambda$. To make our analysis transparent we are dealing here with
the chiral limit again putting the current quark mass equal to zero and further follow the same line as
in the considered KKB model. Then the energy density of ensemble in the phase of non-zero chiral
condensate is given by
\begin{eqnarray}
%28
\label{ncc2}
&&H_{cc}=-2N_c~\int\limits_{P_{\mbox{\scriptsize{F}}}}^{\Lambda}\!\!d \widetilde{\vf p}~
P_0+\frat{M^2}{4g_s}=%\nonumber\\[-.1cm]\\ [-.1cm]
-\frat{N_c}{4\pi^2}~\Biggl[\Lambda E_\Lambda^3-
P_{\mbox{\scriptsize{F}}}E_{\mbox{\scriptsize{F}}}^3-\nonumber\\
&&\hspace{-0.3cm}-\frat{M_q^2}{2}\Bigl(\Lambda E_\Lambda-P_{\mbox{\scriptsize{F}}}
E_{\mbox{\scriptsize{F}}}\Bigr)-\frat{M_q^4}{2}\ln \frat{\Lambda+E_\Lambda}{P_{\mbox{\scriptsize{F}}}+
E_{\mbox{\scriptsize{F}}}}\Biggr]+\frat{N_c}{2G_s}M^2,
%\nonumber
\end{eqnarray}
The bayon number density in this model is given by (\ref{n3}).
%%%%%%%%%%%%%%%%%%%%%%%%%%%%%%%%%%%%%%%%%%%%%%%%%
\begin{figure}%[!tbh]
\includegraphics[width=0.3\textwidth]{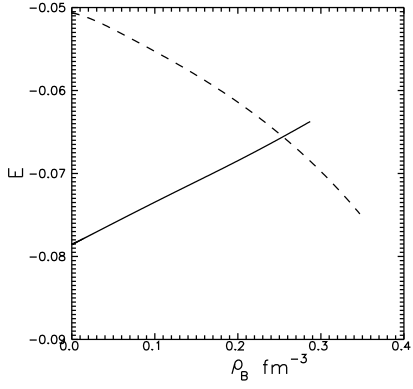}
\caption{The energy of the chiral phase (solid line) and the phase of color
superconductor (dashed curve) as a functions of baryon number density for
the NJL model.
}
\label{f7}
\end{figure}
%%%%%%%%%%%%%%%%%%%%%%%%%%%%%%%%%%%%%%%%%

The gap equation for this model develops the following form
\begin{eqnarray}
%29
\label{p29}
&&\Delta=\Delta_q~\frat{G_d}{4\pi^2}~
\Bigl[(\Lambda-\mu) E_{\Lambda+\mu}+(\Lambda+\mu) E_{\Lambda-\mu}+
\mu E_\mu+\nonumber\\[-.1cm]\\ [-.1cm]
&&+(2\mu^2-\Delta^2)\ln \left(\frat{\Lambda+\mu+E_{\Lambda+\mu}}{\Delta}
\frat{\Lambda-\mu+E_{\Lambda-\mu}}{\Delta}\right)\Bigr]~.
\nonumber
\end{eqnarray}
An artificial introduction of a small bare mass $\Delta_o<<\Delta$ allows us easily to solve
this equation by iterating. Then the contribution of color doublet into the energy density
for the NJL model is given by
\begin{eqnarray}
%30
\label{p30}
&&H_d=-2\cdot 2\int\limits^\Lambda d \widetilde{\vf p} ~\frat12~(P_++P_-)+\frat{\Delta^2}{4g_d}=\\
&&-2\left\{\frat{1}{8\pi^2}\biggl[(\Lambda+\mu)
E_{\Lambda+\mu}^3+(\Lambda-\mu) E_{\Lambda-\mu}^3-\right.\nonumber\\
&&\left.-\frat{\Delta^2}{2}\Bigl((\Lambda+\mu)
E_{\Lambda+\mu}+(\Lambda-\mu) E_{\Lambda-\mu}\biggr)-\right.\nonumber\\
&&\left.-\frat{\Delta^4}{2}\ln \left(
\frat{\Lambda+\mu+E_{\Lambda+\mu}}{\Delta}\frat{\Lambda-\mu+
E_{\Lambda-\mu}}{\Delta}\right)\biggr]+\right.\nonumber\\
&&\left.+\frat{\mu^2}{4\pi^2}\biggl[(\Lambda+\mu) E_{\Lambda+\mu}+
(\Lambda-\mu) E_{\Lambda-\mu}+\right.\nonumber\\
&&\left.+\Delta^2\ln \left(
\frat{\Lambda+\mu+E_{\Lambda+\mu}}{\Delta}\frat{\Lambda-\mu+
E_{\Lambda-\mu}}{\Delta}\right)\biggr]-\right.\nonumber\\
&&\left.-\frat{\mu}{3\pi^2}\biggl[E_{\Lambda+\mu}^3-
E_{\Lambda-\mu}^3+2\Delta^2-2E_\mu^3\biggr]\right\}+
\frat{\Delta^2}{G_d}~.\nonumber
\end{eqnarray}
The baryon number density in the color superconducting phase looks like
\begin{eqnarray}
%31
\label{p31}
&&Q_d=2\int\limits^\Lambda d \widetilde{\vf p}
~\left(\frat{E_p+\mu}{P_+}-\frat{E_p-\mu}{P_-}\right)=\\
&&=2\left\{\frat{1}{6\pi^2}\biggl[E_{\Lambda+\mu}^3-E_{\Lambda-\mu}^3\biggr]+\frat{\mu^2-\Delta^2}{2\pi^2}
\biggl[E_{\Lambda+\mu}-E_{\Lambda-\mu}\biggr]-\right.\nonumber\\
&&\left.-\frat{\mu}{2\pi^2}\biggl[(\Lambda+\mu)E_{\Lambda+\mu}+
(\Lambda-\mu)E_{\Lambda-\mu}-\right.\nonumber\\
&&\left.-\Delta^2\ln \left(
\frat{\Lambda+\mu+E_{\Lambda+\mu}}{\Delta}\frat{\Lambda-\mu+
E_{\Lambda-\mu}}{\Delta}\right)\biggr]\right\}
~.\nonumber
\end{eqnarray}
Now we have the full setup to perform an analysis similar to that done for the KKB model. In particular,
for example, the Fermi momentum dependence on chemical potential $P_{\mbox{\scriptsize{F}}}(\mu)$ (see
Fig. \ref{f4}) occurs almost linear function but we do not present these data here. Fig. \ref{f6}
demonstrates the dynamical quark mass (solid line) and the gap (dashed line) in the NJL model as the
functions of baryon number density. In distinction with the KKB model at low density the gap is not
appreciably developing, although, in fact, it is easy to see from the equation system (\ref{p26}) that
at low values of chemical potential we are dealing with the equation of the same type as an equation for
the dynamical quark mass. The reason why the gap does not reach an appreciable value is related to the
smallness of coupling constant $G_d$ comparing to the constant $G_s$ (it is weaker approximately in factor
three) and the corresponding integral can not provide a proper critical value. It should be noted, however,
that at high baryon number densities the values obtained values are not reliable because the characteristic
quark momenta available in the problem become quite comparable with the cutoff parameter $\Lambda$.
Fig. \ref{f7} demonstrates the energy (in some arbitrary units) of non-zero chiral condensate phase
and the phase of color superconductor (dashed curve) as a functions of baryon number density.
Comparing this figure to the figure \ref{f5} we can see that the phase transition boundary in the color
superconductor state in both models are roughly the same. It is also easy to guess how to rewrite the
necessary formulae for the arbitrary form-factor $F({\vf p})$.

The work was supported by the State Fund for Fundamental Research of Ukraine, Grant \nomer{Ph58/04}.

%\newpage


\begin{thebibliography}{99}
%1
\bibitem{ebw}
G. Endrodi, QCD phase diagram: overview of recent lattice results, arXiv:1311.0648 [hep-lat];\\
S. Borsanyi, Thermodynamics of the QCD transition from lattice, arXiv:1210.6901 [hep-lat];\\
J. Wambach, Recent theoretical developments in the QCD phase diagram, arXiv:1111.5475 [hep-ph].
%2
\bibitem{polch}
J. Polchinski, Effective field theory and the Fermi surface, arXiv:9210046 [hep-th].
%3
\bibitem{blr}
D. Bailin and A. Love, Phys. Rep.  {\bf 107} (1984) 325;\\
M. Alford, K. Rajagopal, F. Wilczek, Phys. Lett. {\bf B422} (1998) 247;\\
J. Berges, K. Rajagopal, Nucl. Phys. {\bf B538} (1999) 215;\\
K. Rajagopal, Nucl. Phys. A {\bf 651} (1999) 150c;\\
R. Rapp, T. Sch\"afer, E. V. Shuryak, and M. Velkovsky,
Ann. Phys. (N.Y.) {\bf 280}  (2000) 35.
%4
\bibitem{dc}
G. W. Carter and D. I. Diakonov, Phys. Rew. D {\bf 60}, 016004 (1999);\\
S. V. Molodtsov and G. M. Zinovjev, Mod. Phys. Lett. {\bf A 18} (2003) 817;\\
G . M. Zinovjev, S. V. Molodtsov, and A. M. Snigirev,
Phys. Atom. Nucl.  {\bf 65} (2002) 929;\\
G. M. Zinovjev and S. V. Molodtsov, Phys. Atom. Nucl.  {\bf 66} (2003) 968; {\bf 66} (2003) 1389.
%5
\bibitem{ker}
A.G. Zubkov, O.V. Dubasov, and B.O. Kerbikov,
Int. J. Mod. Phys. A14 (1999) 241.
%6
\bibitem{ko}
Jun Xu, Taeso Song, Che Ming Ko, and Feng Li, arXiv:1308.1753 [nucl-th].

%7
\bibitem{MZ}
G. M. Zinovjev and S. V. Molodtsov, Theor. Mat. Fiz.  {\bf 160} (2009) 444;\\
S. V. Molodtsov and G. M. Zinovjev, Phys. Rev. {\bf D80} (2009) 076001;\\
S. V. Molodtsov, A. N. Sissakian and G. M. Zinovjev,
Europhys. Lett. {\bf 87} (2009) 61001;\\
S. V. Molodtsov, A. N. Sissakian and G. M. Zinovjev,
Ukr. J. Phys. {\bf 8}--{\bf 9} (2009) 775.
%8
\bibitem{njl}
Y. Nambu and G. Jona-Lasinio, Phys. Rev. {\bf 122} (1961) 345.
%9
\bibitem{kldsh}
M. V. Sadovskii, Diagrammatics, Singapore: World Scientific, 2006.\\
L. V. Keldysh, Doctor thesis, FIAN, (1965);\\
E. V. Kane, Phys. Rev. {\bf 131} (1963) 79;\\
V. L. Bonch-Bruevich, in 'Physics of solid states', M., VINITI, (1965).
%10
\bibitem{5}
T. Hatsuda and T. Kunihiro, Phys. Rep. {\bf 247} (1994) 221.
%11
\bibitem{cor1}
G. S. Bali, N. Brambilla, and A. Vairo, Phys. Lett. B {\bf 421}, 265 (1998);
Y. Koma, M. Koma, Nucl. Phys. B {\bf 769}, 79 (2007).\\
Yu. A. Simonov and V. I. Shevchenko, Adv. High Energy, 2009, 873051 (2009);
arXiv: 0902.1405 [hep-ph];\\
Yu. A. Simonov, arXiv: 1003.3608 [hep-ph].
%12
\bibitem{cor2}
A. Di Giacomo, E. Meggiolaro, and H. Panagopoulos,
Nucl. Phys. B {\bf 483}, 371 (1997);\\
M. D' Elia, A. Di Giacomo, and E. Meggiolaro,
Phys. Lett. B {\bf 408}, 315 (1997);\\
G. Bali, N. Brambilla, and A. Vairo, Phys. Lett. B {\bf 421}, 265 (1998);\\
M. D\'Elia, A. Di Giacomo, and E. Meggiolaro, Phys. Rev. D {\bf 67}, 114504
(2003);\\
A. E. Dorokhov, S. V. Esaibegyan, and S. V. Mikhailov, Phys. Rev. D {\bf 56},
4062 (1997);\\
E.-M. Ilgenfritz, B. V. Martemyanov, S. V. Molodtsov, M. M\"uller-Preussker, and
Yu. A. Simonov, Phys. Rev. D {\bf 58}, 114508 (1998);\\
E.-M. Ilgenfritz, B. V. Martemyanov, and M. M\"uller-Preussker,
Phys. Rev. D {\bf 62},  096004 (2000).
%13
\bibitem{cor3}
E.-M. Ilgenfritz, M. M\"uller-Preussker, A. Sternbeck, A. Schiller,
and I. L. Bogolubsky, Braz.J.Phys. {\bf 37} (2007) 193;\\
I. L. Bogolubsky, V. G. Bornyakov, G. Burgio, E.-M. Ilgenfritz,
M. M\"uller-Preussker, and V. K. Mitrjushkin,
13th Lomonosov Conference on Elementary Particle Physics,
Moscow, August 2007, arXiv: 0804.1250 [hep-lat].
%14
\bibitem{cah}
R. T. Cahill, J. Praschifka,  and C. J. Burden, Aust. J. Phys. {\bf 42} (1989) 161.
%15
\bibitem{MZ2}
G. M. Zinovjev, S. V. Molodtsov, and A. N. Sissakian, Phys. At. Nucl. {\bf 73} (2010) 1245;\\
S. V. Molodtsov and G. M. Zinovjev, Europhys. Lett. {\bf 93} (2011) 11001;\\
S. V. Molodtsov and G. M. Zinovjev, Phys. Rev. {\bf D84} (2011) 036011;\\
G. M. Zinovjev and S. V. Molodtsov, Phys. At. Nucl. {\bf 75} (2012) 262.
%16
\bibitem{MZ3}
G. M. Zinovjev, S. V. Molodtsov,  Phys. of Elemen. Part. and Atomic Nucl. {\bf 44} (2013) 577.
%17
\bibitem{TM}
H. Tezuka, Phys. Rev. {\bf C22} (1980) 2585;  {\bf C24} (1981) 288;\\
G. Baym and S. A. Chin, Nucl. Phys. {\bf A262} (1976) 537;\\
T. Matsui, Nucl. Phys. {\bf A370} (1981) 369.

\end{thebibliography}
\end{document}